\documentclass[12pt]{article}

\usepackage{graphicx}

\newcommand{\be}{\begin{equation}}
\newcommand{\ee}{\end{equation}}
\newcommand{\bea}{\begin{eqnarray}}
\newcommand{\eea}{\end{eqnarray}}

 \makeatletter
\newdimen\normalarrayskip
\newdimen\minarrayskip
\normalarrayskip\baselineskip
\minarrayskip\jot
\newif\ifold             \oldtrue

\makeatother
\newlength{\extraspace}
\newlength{\extraspaces}
\begin{document}

\begin{center}
\baselineskip=24pt

{\bf {\large {Post-measurement Nonlocal Gates}}}\\[5mm]

\baselineskip=12pt

{Daegene Song} \\[3mm]
{\it {National Institute of Standards and Technology,\\ 100 Bureau Drive, MS 8910, Gaithersburg, MD 20899}} \vspace{.5cm}

{\bf Abstract}
\vspace{2mm}

\begin{minipage}{13cm}
\baselineskip-12pt 

Several proposed quantum computer models include measurement processes,
in order to implement nonlocal gates and create necessary entanglement resources during the computation.
We introduce a scheme in which the measurements can be delayed for two- and three-qubit nonlocal
gates.  We also discuss implementing arbitrary nonlocal gates when measurements are included during
the process.

\end{minipage}
\end{center}
\vspace{.5cm}

Various people have introduced \cite{eisert,collins,gottesman} nonlocal gates so that when entanglement resources
are provided, one can perform multiple qubit gates on widely separated qubits using only local operations.
 Using the concept of an entanglement bus, a quantum computer architecture has been
 proposed \cite{brennen}  to efficiently implement nonlocal gates on a lattice model,
allowing only nearest neighbor interactions.
However, these nonlocal gates involve measurement processes during the implementation of the gates.
If they were to be used during a large-scale quantum computation,
a large number of measurements would be required \cite{brennen2}.
This will be particularly difficult for systems where measurement time is substantially longer than unitary operation time or
 the case when measurement errors are greater than errors due to unitary operations.
It is therefore desirable to seek a possibility such that all the measurements can be delayed until the end of the
computation.  As a first step, we introduce a scheme in which measurements
can be delayed until the end of the implementation of two- and three-qubit nonlocal gates.

Eisert {\it {et al.}} \cite {eisert} (also in \cite{collins})
have discussed the use of classical and entanglement resources  for the
implementation of nonlocal gates.  Here, we would like to consider implementing nonlocal
gates on a quantum computer.  Since we have a rather reliable classical information processor,
we will be concerned mainly with entanglement resources rather than classical information.
Firstly, we will review the implementation of nonlocal gates with measurements in the
middle of the process.  Arbitrary quantum gates on two distant qubits can be performed
with two EPR pairs, or ebits, by using teleportation.

\begin{figure}

\includegraphics[scale=.8]{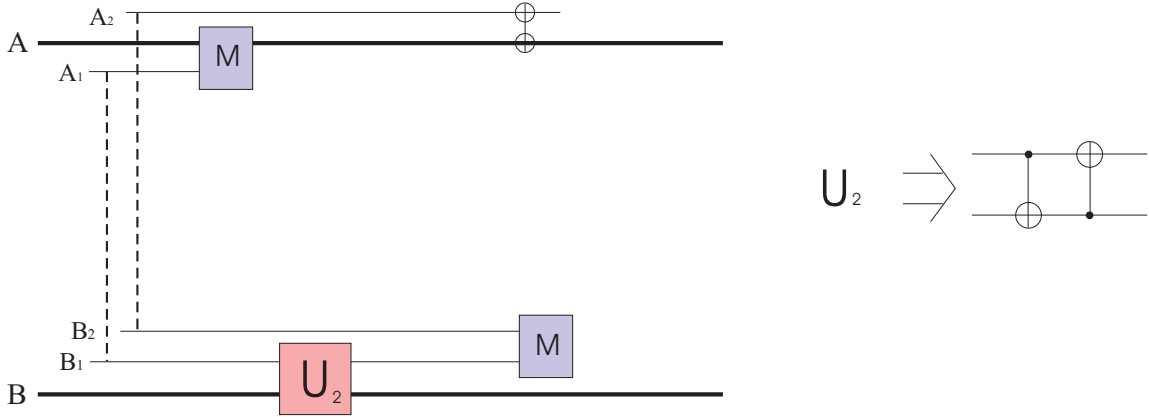}
\caption{Arbitrary nonlocal two qubit gate, $U_2$, between $A$ and $B$ using teleportation. Using ebit $A_1 B_1$, $A$ is teleported to
$B_1$. The desired two qubit gate is performed on $B_1 B$, we then teleport $B_1$ back using ebit $A_2 B_2$ and swap $A_2$ with $A$. $M$ refers to the Bell measurement.   }
\label{before}
\end{figure}

Let us take an example to see how it works. As shown in  Fig. \ref{before}, given two distant localized states $(a|0\rangle + b|1\rangle )_{A}$ and
$(c|0\rangle + d|1\rangle )_B$,  we want to perform nonlocal gates between $A$ and $B$ using
entanglement resources, in this case, we want to apply CNOT with the control qubit $A$ and target $B$ followed by
another CNOT with the control $B$ and target $A$.  We start by teleporting qubit $A$ using an ebit of $A_1$ and $B_1$,
\begin{equation}
(a|0\rangle + b|1\rangle )_{A} (|00\rangle + |11\rangle )_{A_1 B_1}
\end{equation}
by making a Bell measurement on $A$ and $A_1$. We will omit the normalization factor when the coefficients are equal.
Bell measurement can be performed by
applying CNOT followed by Hadamard gate which transforms, $|\phi^+\rangle\equiv |00\rangle + |11\rangle   \rightarrow |00\rangle$,
$|\psi^+\rangle \equiv |01\rangle +|10\rangle \rightarrow |01\rangle$, $|\psi^-\rangle \equiv |01\rangle -|10\rangle \rightarrow |11\rangle$ and $|\phi^-\rangle \equiv |00\rangle - |11\rangle \rightarrow |10\rangle$.
But we will just use the original Bell basis, $|\phi^{\pm}\rangle,|\psi^{\pm}\rangle$, for simplicity.
 Depending on the Bell measurement result on $A A_1$, the following correction gates are applied to $B_1$, $|\phi^+\rangle \rightarrow {\bf {1}}$, $|\psi^+\rangle \rightarrow \sigma_x$,
 $|\psi^-\rangle \rightarrow \sigma_z\sigma_x  $, and $|\phi^-\rangle \rightarrow \sigma_z$, yielding
 the result $(a|0\rangle + b|1\rangle)_{B_1} $.
Next we perform the desired arbitrary local operation on $B_1$ and $B$, i.e. $U_2 \equiv {\rm {CNOT}}_{B B_1 }$ ${\rm {CNOT}}_{B_1 B}$.  The result becomes,
\begin{equation}
\left( ac|00\rangle  + ad|11\rangle + bc|01\rangle + bd|10\rangle\right)_{B_1 B}
\label{tele}\end{equation}
We then use another ebit $(|00\rangle + |11\rangle )_{A_2 B_2}$ and perform Bell measurement
 on $B_1$ and $B_2$.  As before, depending on the results, we apply the same correction gates to $A_2$.
 Finally swapping $A_2$ with $A$ yields the final state we want as in (\ref{tele}) for qubits
 $A$ and $B$.  It is clear this procedure can be generalized to arbitrary two qubit operations for $U_2$.

In a similar manner, this can be generalized to $n$ qubit arbitrary nonlocal gates with
 $2(n-1)$ ebits.  We use $(n-1)$ ebits to teleport $(n-1)$ of the qubits near  to the $n$th qubit, and then perform the desired
 gates on them and teleport each back using another $(n-1)$ ebits.

\begin{figure}
\includegraphics[scale=.8]{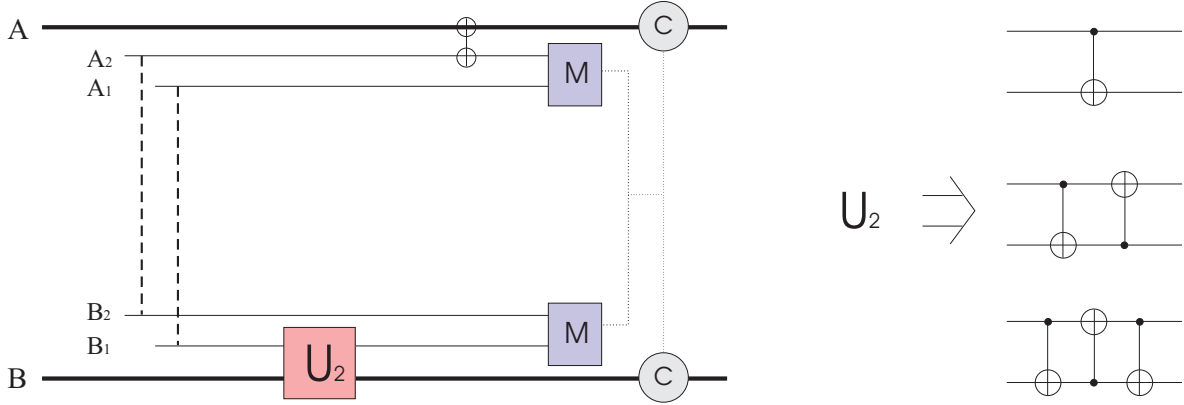}
\caption{Post-measurement two-qubit nonlocal gate for one, two and three CNOT's.  Using two ebits $A_1 B_1$ and $A_2 B_2$, two qubit
gates are performed on $B_1$ and $B$ and $A_2$ and $A$ are swapped.  The measurements are made at the end, and
the correction gates dependent on $U_2$ are applied to $A$ and $B$. }
\label{twoQUBIT}
\end{figure}

 So far, we've considered the nonlocal gates where measurements were involved in the middle of the process as in previous studies \cite{eisert,collins}.
In the following, we introduce two-qubit and three-qubit nonlocal gates where measurements may be delayed until the
end of the gate.  We first consider a nonlocal CNOT gate on qubits $A$ and $B$ using the ebit $A_1 B_1$ as follows,
\begin{equation}
(a|0\rangle + b|1\rangle )_A (|00\rangle + |11\rangle )_{A_1 B_1} (c|0\rangle + d|1\rangle )_B
\end{equation}
This can be re-written as
\begin{eqnarray}
 \{   |\phi^+\rangle_{AA_1} (a|0\rangle + b|1\rangle )_{B_1} &+&  |\psi^+\rangle_{AA_1} (a|1\rangle + b|0\rangle )_{B_1} \nonumber \\
 + |\psi^-\rangle_{AA_1} (a|1\rangle - b|0\rangle )_{B_1} &+&  |\phi^-\rangle_{AA_1} (a|0\rangle - b|1\rangle )_{B_1} \}
      \times    (c|0\rangle + d|1\rangle)_B
\end{eqnarray}
We start by applying the desired CNOT gate on $B_1$ and $B$.  Next we use the second ebit $(|00\rangle + |11\rangle )_{A_2 B_2}$ as shown in Fig. \ref{twoQUBIT}
 and swap $A$ with $A_2$.
It now remains to perform two local Bell measurements on $B_1 B_2$ and $A_1 A_2$.  The result on $B_1 B_2$ will
correct qubit $A$ as $|\phi^+\rangle \rightarrow {\bf {1}}$, $|\psi^+\rangle \rightarrow \sigma_x$,
 $|\psi^-\rangle \rightarrow \sigma_z\sigma_x  $, and $|\phi^-\rangle \rightarrow \sigma_z$ .  Next, the result on $A_1 A_2$ will correct both
qubits $A$ and $B$ as follows, $\phi^+ \rightarrow {\bf {1}}^A \otimes {\bf {1}}^B$, $\psi^+ \rightarrow \sigma_x^A \otimes \sigma_x^B$,
$\psi^- \rightarrow (\sigma_z \sigma_x )^A \otimes \sigma_x^B$, $\phi^- \rightarrow \sigma_z^A \otimes {\bf {1}}^B$.

This procedure can be  generalized to other networks of CNOT gates.
Let us consider a CNOT gate with control qubit $A$ and target $B$ followed by another CNOT with  control $B$ and target $A$, i.e. $U_2 = {\rm {CNOT}}_{BA}{\rm {CNOT}}_{AB}$.
We follow the same network as in Fig. 2 and the result on $B_1 B_2$ will correct just as in a single CNOT gate case.
Then the result on $A_1$ and $A_2$ corrects again both $A$ and $B$ as follows, $\phi^+ \rightarrow {\bf {1}}^A \otimes {\bf {1}}^B$, $\psi^+ \rightarrow {\bf {1}}^A \otimes \sigma_x^B$, $\psi^- \rightarrow \sigma_z^A \otimes (\sigma_z \sigma_x)^B$,
$\phi^- \rightarrow \sigma_z^A \otimes \sigma_z^B$.  For three CNOT gates, i.e. a swap operation between $A$ and $B$, the correction gate for $A_1A_2$ is applied only to
qubit $B$ as follows, $\phi^+ \rightarrow {\bf {1}}, \psi^+ \rightarrow \sigma_x^B$, $\psi^- \rightarrow (\sigma_z\sigma_x )^B$,
$\phi^- \rightarrow \sigma_z^B$.  Therefore, for a swap nonlocal gate, the correction gate for $A_1A_2$ is applied to only
qubit $B$ while the correction gate for $B_1B_2$ is applied to $A$ as before.

\begin{figure}
\includegraphics[scale=.8]{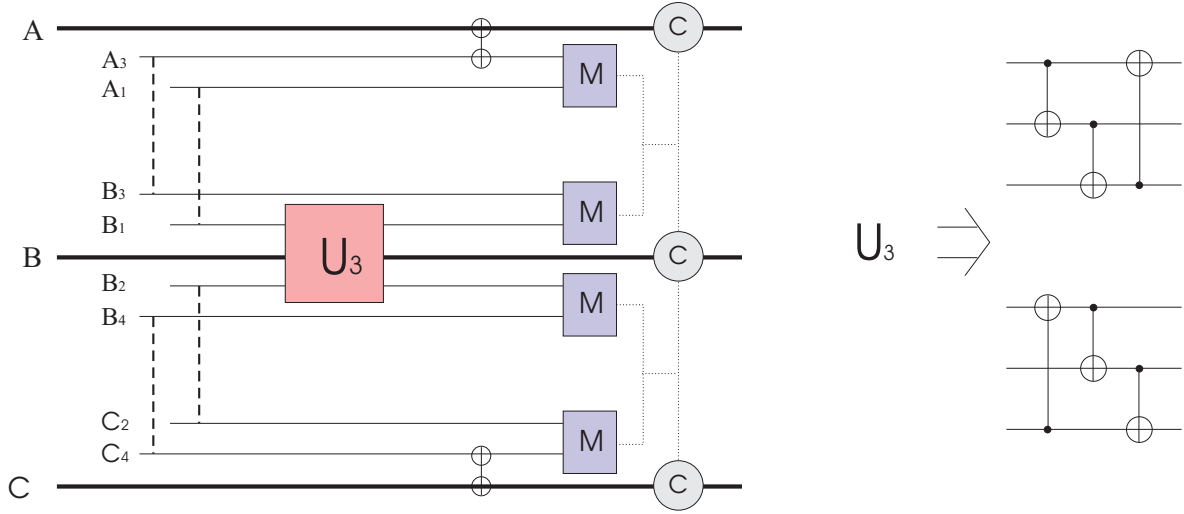}
\caption{Post-measurement nonlocal three-qubit gate.  Here, four ebits are used: $A_1 B_1$, $B_2 C_2$, $A_3 B_3$ and $B_4 C_4$.  }
\label{threeQUBIT}
\end{figure}

Next we consider the case of operations on three qubits.  Three arbitrary qubits $(a|0\rangle + b|1\rangle )_A$, $(c|0\rangle + d|1\rangle )_B$, and $(e|0\rangle + f|1\rangle )_C$ with two ebits $A_1 B_1$ and $B_2 C_2$ can be
written as
\begin{eqnarray}
 \{    |\phi^+\rangle (a|0\rangle + b|1\rangle ) &+&  |\psi^+\rangle (a|1\rangle + b|0\rangle ) \nonumber \\
    |\psi^-\rangle (a|1\rangle - b|0\rangle ) &+&  |\phi^-\rangle (a|0\rangle - b|1\rangle ) \}_{AA_1 B_1} \otimes (c|0\rangle + d|1\rangle )_B  \nonumber \\
& \otimes  & \{ ( e|0\rangle + f|1\rangle )|\phi^+\rangle + (e|1\rangle + f|0\rangle )|\psi^+\rangle \nonumber \\
& + & (e|1\rangle - f|0\rangle )|\psi^-\rangle + (e|0\rangle - f|1\rangle )|\psi^+\rangle \}_{B_2 CC_2}
\end{eqnarray}
First we study the network $U_3={\rm {CNOT}}_{CA}{\rm {CNOT}}_{BC}{\rm {CNOT}}_{AB}$ as shown in Fig. \ref{threeQUBIT}.  As in the two-qubit case, we apply
these three CNOTs on $B_1,B$ and $B_2$, i.e. $U_3\equiv{\rm {CNOT}}_{B_2 B_1}{\rm {CNOT}}_{BB_2}{\rm {CNOT}}_{B_1 B}$.
Then using two additional ebits $A_3 C_3$ and $B_4 C_4$, we swap $A_3$ with $A$ and $C_4$ with $C$.
It then suffices to make Bell measurements on $A_1 A_3$, $B_1B_3$, $B_2B_4$ and $C_2C_4$.
For correction gates, $B_1 B_3$ will correct $A$ and $C_2 C_4$ will correct $C$ as usual, $|\phi^+\rangle \rightarrow {\bf {1}}$, $|\psi^+\rangle \rightarrow \sigma_x$,
 $|\psi^-\rangle \rightarrow \sigma_z\sigma_x  $, and $|\phi^-\rangle \rightarrow \sigma_z$, respectively.
After these corrections, we then correct $A,B$ and $C$ with the result on $A_1A_3$ and $C_2 C_4$ as shown in the Table \ref{corr3}.

We consider another three-qubit nonlocal gate with $U_3 \equiv {\rm {CNOT}}_{BC}{\rm {CNOT}}_{AB}{\rm {CNOT}}_{CA}$.  We follow the same procedure as
before and the measurement result on $B_1B_3$ and $B_2B_4$ correct $A$ and $C$ respectively as before.
The correction gate for $A_1A_3$ and $B_2B_4$ is shown in the Table \ref{correct}.

\begin{table*}

\begin{center}
\begin{tabular}{|l||l|l|l|l|} \hline
 & $|\phi^+\rangle_{C_2C_4}$ & $|\psi^+\rangle_{C_2C_4}$ & $|\psi^-\rangle_{C_2C_4}$ & $|\phi^-\rangle_{C_2C_4}$ \\ \hline\hline
$|\phi^+\rangle_{A_1 A_3}$  &  ${\bf {1}}\otimes {\bf {1}} \otimes {\bf {1}}$                 & $\sigma_x \otimes {\bf {1}} \otimes \sigma_x $                         & $\sigma_x \otimes \sigma_z \otimes \sigma_z \sigma_x $              & ${\bf {1}} \otimes \sigma_z \otimes \sigma_z $   \\
$|\psi^+\rangle_{A_1 A_3}$  &  ${\bf {1}} \otimes \sigma_x \otimes \sigma_x $           & $ \sigma_x \otimes \sigma_x \otimes {\bf {1}} $                        & $\sigma_x \otimes \sigma_z\sigma_x \otimes \sigma_z $               &  ${\bf {1}} \otimes \sigma_z\sigma_x \otimes \sigma_z\sigma_x $ \\
 $|\psi^-\rangle_{A_1 A_3}$ &  $\sigma_z \otimes \sigma_x \otimes \sigma_z\sigma_x $  & $\sigma_z\sigma_x \otimes \sigma_x \otimes \sigma_z $                & $\sigma_z\sigma_x \otimes \sigma_z\sigma_x \otimes {\bf {1}}  $   & $\sigma_z \otimes \sigma_z\sigma_x \otimes \sigma_x $\\
 $|\phi^-\rangle_{A_1 A_3}$ &  $\sigma_z \otimes {\bf {1}} \otimes \sigma_z $           & $\sigma_z\sigma_x \otimes {\bf {1}} \otimes \sigma_z\sigma_x $     & $\sigma_z\sigma_x \otimes \sigma_z \otimes \sigma_x  $              & $\sigma_z \otimes \sigma_z \otimes {\bf {1}} $\\  \hline
\end{tabular}

\end{center}
\caption{Correction gates based on Bell measurements on $A_1 A_3$ and $C_2 C_4$. They are applied to qubits $A$,$B$ and $C$ for the post-measurement nonlocal three-qubit gate in the Fig. \ref{threeQUBIT}
network for the case of ${\rm {CNOT}}_{CA} {\rm {CNOT}}_{BC} {\rm {CNOT}}_{AB}$.} \label{corr3}\end{table*}

\begin{table*}

\begin{center}
\begin{tabular}{|l||l|l|l|l|} \hline
 & $|\phi^+\rangle_{C_2C_4}$ & $|\psi^+\rangle_{C_2C_4}$ & $|\psi^-\rangle_{C_2C_4}$ & $|\phi^-\rangle_{C_2C_4}$ \\ \hline\hline
$|\phi^+\rangle_{A_1 A_3}$  &  ${\bf {1}}\otimes {\bf {1}} \otimes {\bf {1}}$                 & $\sigma_x \otimes \sigma_x \otimes {\bf {1}} $                         & $\sigma_x \otimes \sigma_z\sigma_x  \otimes \sigma_z $              & ${\bf {1}} \otimes \sigma_z \otimes \sigma_z $   \\
$|\psi^+\rangle_{A_1 A_3}$  &  $\sigma_x \otimes \sigma_x \otimes \sigma_x $           & $ {\bf {1}} \otimes {\bf {1}} \otimes \sigma_x $                        & ${\bf {1}} \otimes \sigma_z \otimes \sigma_z\sigma_x $               &  $\sigma_x \otimes \sigma_z\sigma_x \otimes \sigma_z\sigma_x $ \\
 $|\psi^-\rangle_{A_1 A_3}$ &  $\sigma_z\sigma_x \otimes \sigma_z\sigma_x \otimes \sigma_z\sigma_x $  & $\sigma_z \otimes \sigma_z \otimes \sigma_z\sigma_x $                & $\sigma_z \otimes {\bf {1}} \otimes \sigma_x  $   & $\sigma_z \sigma_x \otimes \sigma_x \otimes \sigma_x $\\
 $|\phi^-\rangle_{A_1 A_3}$ &  $\sigma_z \otimes \sigma_z \otimes \sigma_z $           & $\sigma_z\sigma_x \otimes \sigma_z\sigma_x \otimes \sigma_z $     & $\sigma_z\sigma_x \otimes \sigma_x \otimes {\bf {1}}  $              & $\sigma_z \otimes {\bf {1}} \otimes {\bf {1}} $\\  \hline
\end{tabular}

\end{center}
\caption{Correction gates based on Bell measurements on $A_1 A_3$ and $C_2 C_4$ for $U_3 \equiv {\rm {CNOT}}_{BC}{\rm {CNOT}}_{AB}{\rm {CNOT}}_{CA}$}
\label{correct}\end{table*}

We have studied some particular networks of CNOT gates for two and three qubits where the measurements
were delayed until the end of the gates.  Further study on the inclusion of other CNOT gates as well as single-qubit operations in the network is
desirable.  Moreover, it is as yet unclear whether the measurements
can be delayed until the end of not only the gates, but also the whole computation.

\end{document}